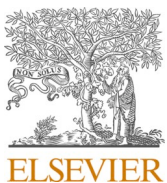
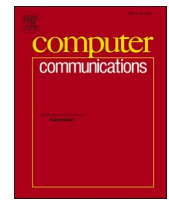

# Enhancing healthcare infrastructure resilience through agent-based simulation methods

David Carramiñana *, Ana M. Bernardos, Juan A. Besada, José R. Casar

*Information Processing and Telecommunications Center, Universidad Politécnica de Madrid, 28040, Madrid, Spain*

A R T I C L E   I N F O

A B S T R A C T

Critical infrastructures face demanding challenges due to natural and human-generated threats, such as pandemics, workforce shortages or cyber-attacks, which might severely compromise service quality. To improve system resilience, decision-makers would need intelligent tools for quick and efficient resource allocation. This article explores an agent-based simulation model that intends to capture a part of the complexity of critical infrastructure systems, particularly considering the interdependencies of healthcare systems with information and telecommunication systems. Such a model enables to implement a simulation-based optimization approach in which the exposure of critical systems to risks is evaluated, while comparing the mitigation effects of multiple tactical and strategical decision alternatives to enhance their resilience. The proposed model is designed to be parameterizable, to enable adapting it to risk scenarios with different severity, and it facilitates the compilation of relevant performance indicators enabling monitoring at both agent level and system level. To validate the agent-based model, a literature-supported methodology has been used to perform cross-validation, sensitivity analysis and test the usefulness of the proposed model through a use case. The use case analyzes the impact of a concurrent pandemic and a cyber-attack on a hospital and compares different resiliency-enhancing countermeasures using contingency tables. Overall, the use case illustrates the feasibility and versatility of the proposed approach to enhance resiliency.

## 1. Introduction

Health Systems (HS) are crucial to ensure the well-being of citizens in modern societies and thus can be considered as a **Critical Infrastructure** (CI). A CI is a system whose facilities, resources, services and entities (e.g.: workers) are vital for the functioning of society and economy [1]. Each CI is usually considered as a complex system composed by interdependent facilities operating as a network to transform resources and provide services to themselves, to other CI sectors or directly to society [2]. In this sense, HS are composed of facilities such as health centres, clinics or hospitals that work as a network referring patients among them. Simultaneously it relies on other CIs such as: the information and telecommunication system (IT system or cyber system, understood as a network of computation and telecommunication nodes), the energy sector, the transportation sector etc; to provide a vital service for society.

However, in recent (and past) years, various natural and human-generated threats have subjected them to stressful situations that have compromised their service quality. For example, the SARS-CoV-2 pandemic has posed an unprecedented challenge, overwhelming hospital care services for extended periods and diverting attention and resources from non-urgent conditions [3]. Furthermore, systematic cyber-attacks suffered by some hospitals have presented an additional obstacle to the proper delivery of patient's healthcare [4,5]. Thus, HS were and are confronted to multiple risks, of different nature, that can arise concurrently. The consequences can range from the immediate impact on citizens' health to significant economic effects, such as those due to increased hours of missed work and rising treatment challenges and costs. Therefore, it seems useful and even necessary to enhance the resilience of HS to better respond to these and other similar stressful situations [6].

As [7] state, "a system is resilient if it can adjust its functioning before, during, or following events (changes, disturbances, or opportunities) and thereby sustain required operations under both expected and unexpected conditions". Thus, in practice, the key principles of resilience engineering [8] include anticipation (risk identification before failures appear), monitoring (continuous control of system's performance and deviation detection), response (strategy development to respond effectively to disruptions), recovery (implementation of measures to recover and restore the system) and learning (analyse incidents and near-misses to identify opportunities for enhancing resilience). This approach has found extensive application in various sectors including aviation, nuclear power generation, or transportation. However, resiliency enhancement is a challenging problem for decision makers, as the operation of CIs is itself complex and heavily relies on interdependencies with other CIs. This makes their evolution and response very difficult to predict [9].

As a possible solution, decision-makers (DM) can benefit from modern **decision support systems** (DSS) that allow modelling and predicting the behaviour of critical systems so that they can anticipate risks and response to them adequately (both in terms of costs and efficacity). Particularly, in the application of resiliency and decision-management tools to healthcare, [10] state the scarcity of comprehensive studies aimed at formulating holistic resilience frameworks capable of incorporating: 1) identification of potential disruptive events, 2) prediction of healthcare system damage, 3) assessment of healthcare system fragility, 4) development of surge and patient demand models, and 5) establishment of functionality, restoration, and interdependence models.

In this regard, this article proposes the usage of an agent-based modelling (ABM) simulation tool as a DSS to bridge the existing gap in the healthcare domain. This contribution aims to evaluate the HS's exposure and damages caused by different concurrent risks and to assess various decision alternatives to restore its functionality. Based on the proposed agent simulation model, the DM should be able to establish a "simulation-based optimization" procedure where different decision scenarios can be formulated, modelled, and compared for their impact on resilience using the model's outputs. Furthermore, to account for interdependencies with other systems and to deal with the existence of compounded mixed risks and emergent behaviours at the HS's level, the integration of the IT system within the simulation model is completed.

The proposal extends our previous work on the 2023 ICT-DM conference [11] with the following refinements.

- It incorporates an enhanced disease evolution model based on a probabilistic Markov chain to enable a fine-grained parameterization of a disease.
- It improves healthcare modelling by introducing additional healthcare services (i.e.: mHealth and in-person consultation) and a queue-based model.
- It explores new metrics to assess the impact of contingency scenarios, including patient waiting times, patient attention times and services utilization ratios.
- It presents an extended validation use-case considering the usage of the proposed DSS tool for both strategic and tactical decision-making.

Overall, the outcome is a **novel HS simulation framework considering IT interdependencies that can be used by DM as a support tool to enhance HS resiliency**, which is validated in a collection of specific crisis scenarios. This proposed model enables obtaining interrelated performance measures both at the lower levels of granularity (services/patients) and at the HS level. The system is open and scalable, and relations with other supporting systems (energy, supplies, …) might be included in the future following a similar approach.

To explain and validate the proposed system, the paper is articulated as follows. First, Section 2 explores the state of the art and compares the proposal against existing literature. Then, the proposed DSS ABM simulator is explained in Section 3. Validation is carried out in Section 4 where both the methodology and the results are laid out. Finally, Section 5 discusses the article conclusions and future work.

## 2. State-of-the-art

This section reviews existing literature addressing decision support systems in healthcare to position the proposed system in the literature landscape. To this end, subsection 2.1 discusses current proposals in terms of their application (i.e.: disease, operational and strategic models) and from the modelling approach point of view (i.e.: analytical and numeric models). It is concluded that there is a lack of DSS for healthcare and it recognizes ABM as a possible modelling approach to cover this gap. In this sense, subsection 2.2 deepens the previous analysis by exploring the applications of ABM to healthcare. Although the analysis shows that ABM is a suitable approach, it also identifies a lack of decision tools at the strategic level. On subsection 2.3, the review analysis is broadened to consider a comprehensive view on HS resilience. Thus, literature that considers HS jointly with other interdependent CIs is explored, finding that interdependences with IT systems are not sufficiently covered. Finally, subsection 2.4 summarizes the previous analyses by performing a comparison of the main advantages and drawbacks of each proposal. As a result of this comparative analysis, it is demonstrated that the proposed system covers an existing knowledge gap.

### 2.1. Decision support systems to enhance healthcare resiliency

As of now, two distinct approaches have been broadly proposed to enhance resilience performance: reflection, involving contemplation and discussion of current practices and ideas, and simulation, entailing the imitation, practice, and rehearsal of real-life events [12]. As [13] states, over the past decade, research on simulation applications have expanded beyond manufacturing environments to include socio-technical systems like healthcare, demonstrating proof-of-concept models for near-future resource planning. Particularly, healthcare structures can take advantage of using simulation techniques for management and planning purposes [14]. In this sense [15], argues that simulation techniques can help DM in the search of the higher efficiency needed within the healthcare sector due to high financial costs.

However [16], highlights the absence of practical frameworks for healthcare resilience and presents one that combines theoretical resilience concepts with heuristic approaches. Similarly [17], has carried out a review of mitigation strategies during the SARS-CoV-2 pandemic and also agrees on the notable scarcity of practical implementations of emerging tools for effectively managing disruptions. They underscore an emphasis on resilience planning, leveraging technologies such as artificial intelligence, blockchain, big data analytics, and simulation.

Tools scarcity might be caused by the complexity of HS which makes them difficult to model, requiring a delicate trade-off between a detailed specific model and a general-purpose model. In this regard, authors of [18] discuss on three different levels of HS models to meet different purposes: **disease models** (used to study particular interventions), **operational models** (used to model the flow of patients within a given hospital department and to provide capacity-demand estimates) and **strategic models** (used to take system-wide investment, long-term decisions).

Due to this variety of purposes, a wide range of quantitative model types has been utilized to examine health services ([19,20]), each suited for analysing specific objectives and facing unique challenges. Thus, from the modelling tool point of view, a distinction can be made between analytical models and numeric simulation models.

On the one hand, **analytical models** are typically employed to tackle problems that can be mathematically well-defined, have a clear problem structure (i.e.: reduced number of variables) and are not affected by stochastic elements. They encompass techniques such as optimization, Markov modelling, Queuing Theory or Optimization methods (e.g.: metamodeling). For example, at an operational level, simulation-based metamodels, or surrogate models, have been previously used in literature to emergency health care, for resource planning, ambulance location, and policy assessment [21]. Metamodeling refers to simplified models that capture the relationship between inputs and outputs of a real system, thus providing a simple and interpretable model for decision-makers. As a drawback, they are constructed based on data or mathematical equations derived from the original system's behaviour, therefore requiring extensive datasets that can be time-consuming to retrieve or generate (from simulation).

Contrary, **numeric simulation** models are more adaptable, flexible and allow to create detailed complex models. However, these advantages come at the expense of an increased computation time and some loss of explainability. Examples of numeric techniques include Montecarlo Simulation (MS), Discrete Event Simulation (DES) models, System Dynamics (SD), and Agent-Based Simulation. Specifically, MS, DES and SD techniques have been concurrently used to model hospitals workflow in Ref. [22]. There, a differential-equation based model is used to predict hospital capacity requirements within the SARS-CoV-2 pandemic. However, authors in Refs. [18,23] argue that those techniques are not well suited to model complex systems, such as those in the healthcare sector. In this sense, they argue that HS procedures involve different alternative processes that can be activated over time based on the decisions on individual actors. This complexity cannot be easily captured using a top-down approach such as SD. Alternatively, they propose the use of a bottom-up technique such as **Agent-Based Models** (ABM) to tackle this problem.

*2.2. Agent based modelling as a decision support system*

Focusing on ABMs, this technique consists in representing a system by the interaction of many autonomous, in general simple agents, that can encompass stochastic components [24]. These agents can represent persons, organizations, or even abstract entities, depending on the context of the modelled system. Agents are modelled individually, then, the macroscopic behaviour of the whole system emerges from the dynamics and interactions between those agents. This inherent agent – system duality of ABMs makes them especially suitable to capture both the micro and macro aspects of complex systems. In this sense, ABM covers a growing need for hybrid simulation models capable of capturing both micro and macro aspects of the socio-economic systems, as stated in Ref. [25]. Moreover, ABM eases the modelling process, as dependencies among systems are included as reasonably well-known, easily explainable, and implementable relationships between agents. Also, once a simulation framework (i.e.: types of agents, considered relationships …) has been established, each agent within the simulation can be modelled with a different level of detail. In fact, the model can also be easily extended with the inclusion of new agents of the same or different nature.

Due to these advantages, ABMs for simulation purposes have been used increasingly within healthcare environments in the last decade to provide disease, operational or strategic level decision-support tools. Starting with the disease-level modelling, a review of simulation technologies used for SARS-CoV-2 research [25] reveals that ABM has been the most frequently used approach (74 % of the considered articles were using this technology), followed by system dynamics model (14 %), discrete simulation (9 %) and, finally, hybrid models (3 %). For instance, in Ref. [26] agent-based modelling capturing individual and social network variations has been employed to assess SARS-Cov-2 spread in the presence of Non Pharmaceutical Interventions (NPI) such as face masks. The performance of ABM has been compared with deterministic modelling. Findings recommend agent-based modelling, as authors suggest that it provides more realistic solutions, incorporating adaptable control measures, while deterministic models tend towards consistent policy enforcement with subsequent relaxation. Similarly, an agent-based simulation model is used in Ref. [27] for the analysis of transmission of SARS-CoV-2 in transit buses under different policies implemented by transit agencies during the early stages of the pandemic in US (e.g. use of KN-95 face masks, open window policies, seating capacity controls, etc.). Also modelling a disease [28], proposes a spatially explicit agent-based influenza model tailored for assessing and advising on influenza control measures, with a focus on Shenzhen city (China) and considering heterogeneous population groups. The three previous cases showcase the effectiveness of ABM to successfully model diseases and take decisions on how to control them. However, they do not assess the impact of those diseases within the HS (e.g.: hospital workload).

Focusing on hospital settings, Ref. [29] reviews the existence of operational models to assess patient flow in emergency departments or to provide insights in facilities design (e.g.: HVAC role in disease spread or pedestrian traffic within a ward). An important caveat of ABM models is highlighted by the authors: ABM models' calibration and validation is not straightforward due to the high number of involved input parameters, emergent behaviours and analysable outputs. Finally, literature also shows an example at a strategic/system level. This is the case of [30] where resource planning strategies are analysed to improve the response of healthcare systems (i.e.: a network of hospitals) under pressure. Considered strategies include the allocation of integrated healthcare resources and patients' transfer, assuming that there are different types of patients and resources to provide access to patient care. Strategies optimization rely on an agent-based continuous-time stochastic model to simulate the parameters that are uncertain. Overall, ABM emerges as a valuable simulation technique whose use in HS has been mostly focused on disease modelling.

*2.3. Modelling healthcare system interdependencies*

Literature analysed so far usually disregards the fact that healthcare facilities do not operate isolatedly but depend on other facilities within the HS. In this sense, HS resiliency can be affected by second-order effects (known as cascading failures): a failure in a healthcare facility can propagate over the healthcare network and impact other hospitals. In this direction, [31] proposes a network-based simulation framework to assess the robustness of a healthcare system accessibility, considering potential cascading failures. To do so, weighted complex networks model patient transfer between nodes, incorporating cascade failure mechanisms to evaluate system robustness under various threat strategies. Results identify vulnerable nodes in healthcare accessibility networks, with a robustness metric combining network efficiency and component size. However, network-based simulation is not suitable for representing micro-level behaviours of isolated entities.

Further exploring network effects, interdependencies with other CI systems (e.g.: electrical grid, water distribution, information systems …) have not been widely studied in literature, as shown in the examples of the previous subsections. However, they cannot be neglected, especially in the case of disasters: failures in other CI might have cascading effects within the HS which might be potentially very relevant. For instance, authors of [9] address these interdependences by analysing how flood-created failures in the electric, water distribution and transportation networks affect a HS.

Another vital interdependence is the one with information and communication systems (more so as the digitalization effort progresses). Notwithstanding, even if information and communication systems simulation has been widely studied [32], a comprehensive practical simulation model has not been previously proposed yet for considering the effect of cyberattacks in a hospital network. In Ref. [33], a partial analysis is performed by evaluating the impact of a ransomware attack on emergency hospital services, from the onset to the recovery phase. A comparison of recovery strategies that includes paying ransom to the attackers, followed by restoration, versus in-house full system restoration from backup are analysed by using a discrete-event simulation model of a typical U.S. urban tertiary hospital. However, the proposal does not consider the existence of compounding threats within different CIs (i.e.: threats that can be simultaneously materialized in interdependent CIs creating synergic negative effects.

## 2.4. Comparison with previous work

The previous analysis has been summarized in Table 1 depicting the following information for each of the reviewed articles: the model detail level, the modelling technique, whether the model considers interdependences and a summary of the positive and negative points. From it, it can be concluded that ABM is a suitable technique to model the complexity of HSs and provide realistic results at multiple resolution levels. Other options such as network-based or SD simulation do not allow modelling HSs with the required flexibility (i.e.: network-based simulation impose representing the HS as a graph) and granularity (i. e.: SD impose a system view). Also, it shows that although there are multiple tools solely focused on helping decision-makers study and prevent disease transmission, there are not as many strategic level tools that are comprehensive (i.e.: model disease transmission, patient severity, different health services …) to allow hospital management to size healthcare resources. Moreover, most of the tools consider healthcare systems in isolation without addressing interdependencies with other CI. This is true for dependencies with IT systems that are only covered once in Ref. [33], without taking advantage of agent-based simulation to facilitate modelling and provide multi-level metrics. In addition, this example ignores the effect of compounding threats that may impact healthcare and IT systems concurrently and are relevant due to possible synergic effects.

In order to cover these gaps (i.e.: strategic tool, IT interdependences and assessing compounding threats), this article leverages the advantages of ABM to go beyond the state-of-the-art by proposing a model that jointly addresses HS and IT simulation, providing useful multi-level metrics to decision makers to enhance HS resiliency.

## 3. Agent based model of a healthcare system

### 3.1. Overall approach

To overcome the shortcomings found in the literature review, this paper proposes a simulation framework based on an ABM that enables healthcare resiliency planning while considering IT interdependencies. In this sense, the HS and IT are abstracted as a CI that is modelled following a bottom-up approach, using a multi-agent system. Within the model, each relevant operation is represented as a set of agents and relationships between them. In ABM, an agent represents an autonomous entity (usually modelling a real facility or actor in the modelled process) that has an internal state and interacts with other agents to build the overall system behaviour by aggregation [24]. Some interactions are modelled as probabilistic/stochastic relations/processes so that statistically relevant realistic outputs are obtained.

The proposed model intends to integrate the complete HS, with patients, disease typologies and care units (e.g. hospitals) with the critical information and communication assets that enable its operation, which can be compromised, seriously affecting HS performance. As a result, the simulation returns a set of timeseries of key metrics about systems performance. These metrics should be analysed by the decision maker to assess the resiliency of the system, as summarized in Fig. 1 (where further details on the ABM are also depicted, as it is decomposed in two main parts, one mainly related to HS modelling and interaction with patients, and another one focused in the IT support to HS).

Besides, this approach enables the examination of various "what-if" scenarios by manipulating the rules and parameters governing agent's behaviour, enabling a simulation-based resiliency optimization approach. Each decision alternative can be easily implemented within the model using parameterizable agents and different simulation scenarios can be configured via a set of configuration parameters. Moreover, the model allows simulating the system both in nominal and non-nominal conditions, considering the existence of contingencies or threats to the system. Particularly, the following contingencies can be easily simulated: 1) the issue and consequences of the propagation of contagious diseases across a population (e.g. SARS-CoV-2); 2) the occurrence of multiple victim incidents for any unexpected reasons; and/or 3) the service degradation due to suffering a cyberattack. Each contingency can be individually modelled, and its occurrence configured in time so that compound and cascading events are possible. To assess the validity of our approach, a high-level and simple behaviour logic encompassing the main trends and dependencies are integrated in the system view.

**Table 1**
Comparison of analysed literature on HS simulators for decision support.

| Reference | Model detail level | Modelling technique | Interdependences | Advantages (+)/Drawbacks (−) |
|---|---|---|---|---|
| [21] | Operational model | Metamodeling | Not considered | +Mathematical simplicity, interpretably and computational speed.<br>- Requires training data and model fitting which can become computational expensive.<br>- Understudied: no general approach to decide which model to use. |
| [22] | Disease model | System dynamics + Montecarlo Simulation + Discrete Event Simulation | Not considered | - SD requires capturing overall system complexity into a single set of interrelated equations.<br>+ DES captures time evolution.<br>+ MS captures uncertainty in the system. |
| [26] | Disease model | Agent Based Simulation | Not considered. | + Realistic results with ABM approach<br>+ It considers Non-Pharmaceutical Interventions (NPI).<br>- The effect on the HS is not considered. |
| [28] | Disease model | Agent Based Simulation | Not considered. | + It considers the effects of NPI.<br>+ It considers demographic attributes of the population.<br>- The effect on the HS is not considered. |
| [27] | Disease model | Agent Based Simulation | Not considered. | + It considers the effects of NPI.<br>- Only micro level is considered.<br>- The effect on the HS is not considered. |
| [29] | Operational model | Agent Based Simulation | Not considered. | - ABM requires difficult validation and verification techniques. |
| [30] | Strategic level | Agent Based simulation | Not considered. | - Patient severity is not considered within the study.<br>- Interdependences are not considered |
| [31] | Strategic model | Network-based simulation | Internal effects (cascading failures). | - Difficult to model<br>- External interdependencies not considered.<br>- Only macro level information is provided. |
| [9] | Strategic model | Network-based simulation | External effects (CI interdependencies with water and electricity CI) | - IT systems not considered |
| [33] | Strategic model | Discrete Event Simulation | External effects (CI interdependencies with IT systems) | + IT systems considered<br>- Compounding threats (e.g.: disease spread) are not considered. |

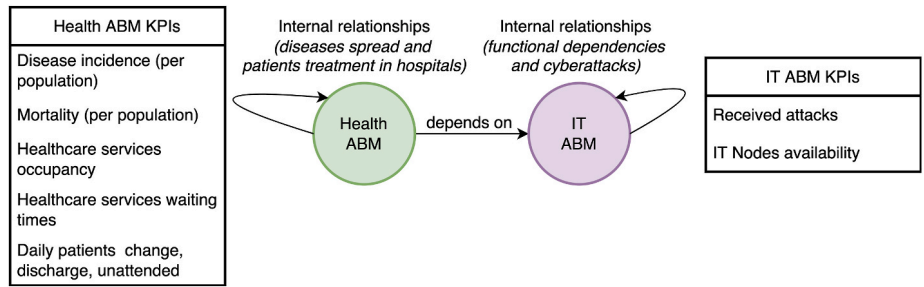

**Fig. 1.** Systemic view of the proposed ABM level. Two ABM systems are modelled with independent internal interactions while enforcing also dependencies between both systems. As a result, a set of KPIs is produced.

*3.2. Healthcare system simulation*

As depicted in Fig. 2, the Healthcare System is modelled as a network of hospitals and populations of potential patients. Three types of agents are considered within the HS model.

- **Population** agents aggregate the healthcare needs of different communities. Therefore, they simulate disease generation and propagation to model patient demand.
- **Patient** agents model the health impact of each disease at an individual level. As the disease evolves through different stages following a stochastical model, patients may require medical treatment in different healthcare levels/services, whose availability and quality of service influences the patient outcome.
- **Hospital** agents model healthcare providing facilities that attend incoming patients subject to capacity restrictions.

Starting with the description of **Population** agents, they dynamically generate new Patient agents following three processes: 1) baseline healthcare demand generation, 2) infectious disease propagation and 3) multiple casualty incidents. Initially, a reference healthcare demand is modelled, to represent the usual healthcare needs of a population (e.g.: chronic diseases, cardiovascular diseases, etc.). Patients are generated over time given a *baseline incidence* (number of affected patients per 100.000 inhabitants). Then, a Susceptible-Infectious-Recovered (SIR) [34] model is used to simulate the spread of contagious diseases within a population and across populations. SIR models simulate infections and recoveries using a set of differential equations parameterized with the *transmission rate* (number of new infections per infected person and time step) and the *recovery rate* (number of recoveries per infected person and time step) of the disease. They usually assume a constant population size and that recovered individuals develop lasting immunity. Finally, multiple casualty incidents can also be included as a sudden surge in healthcare demand. A combination of these three processes can be simultaneously used to generate realistic demand patterns.

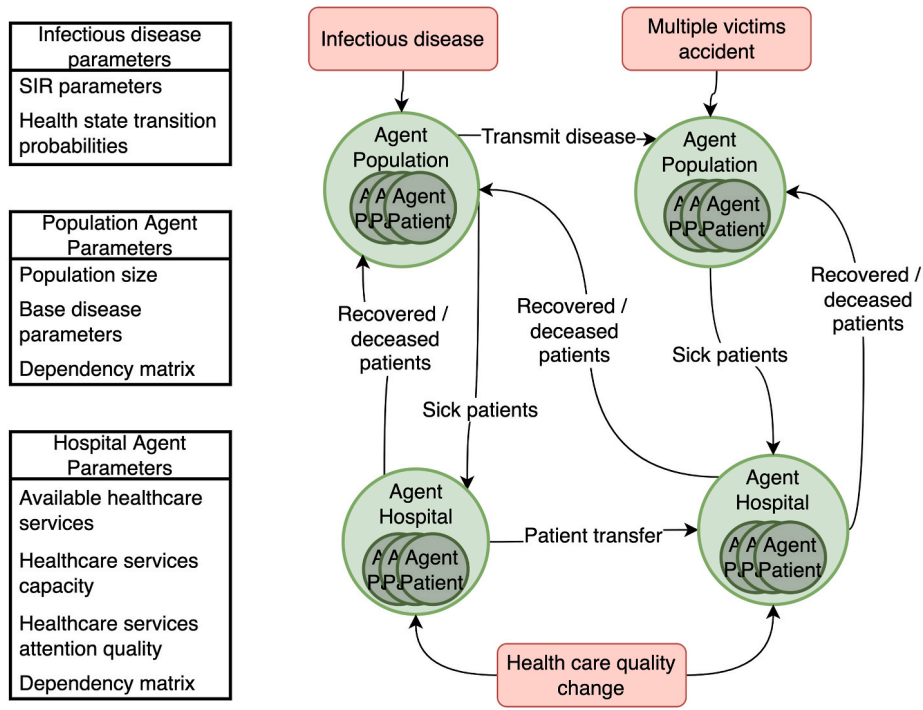

**Fig. 2.** ABM model of the healthcare system.

Once a patient contracts a disease, it is proposed to statistically model its evolution at an individual level (**Patient** agent) using a Markov chain with different health states (defined by set $H$), as is common in literature [35]. Generated ill patients can traverse through 5 different illness states: very mild symptoms, mild symptoms, moderate symptoms, severe symptoms, and critical symptoms; before recovering or eventually dying from the disease, as shown below:

$h \in \{healthy, very\ mild, mild, moderate, severe, critical, death\} \equiv H$

From all the possible transitions within a Markov model of 7 states, transitions are restricted (as shown in Fig. 3 and defined in set T) to those that imply recovering to the healthy state, staying in the same health state, worsening to the next health state or dying:

$t \in \{stay, worsening, recovery, death\} \equiv T$

In Markov models, transitions between states are governed by stationary probabilities, which will be denoted $p(t \mid h)$, meaning the probability of performing transition $t$ from state $h$. Given a state, the sum of all its transition probabilities must equal to 1:

$p(stay|h) + p(worsening|h) + p(recovery|h) + p(death|h) = 1$

Then, within the simulation loop, random experiments following the transition probabilities distributions of a given state are used to simulate changes of states.

As the proposed Markov model aims to imitate a disease progression, transition probabilities must be crafted to imitate it in terms of the length of stay at each state, worsening rates or death rates. Literature shows that transition probabilities can be obtained from aggregate administrative data present in Electronic Health Records, which would be readily available for HS decision-makers [36]. One of such readily available information is the average length of stay in each of the health states ($h \in H$) that we will denote as follows: $T_h$.

It seems logic to assume that the number of time steps required for a health state change follows a geometric distribution of mean $T_h$. This means that, as time passes, it is expected that the patient will either recovery or worsen, decreasing the probability of continuing in the same state. Given this assumption and due to the properties of the geometric distribution, the probability of changing of state at any time step is $\frac{1}{T_h}$. Then, the transition probability of staying in the same state can be computed as:

$$P(stay \mid h) = 1 - \frac{1}{T_h}$$

The rest of transition probabilities can be obtained from aggregated statistics describing the overall prognosis of a given disease under study. This means statistics indicating the proportion of patients that recover from a given state or that die from a disease level. Again, it is expected that this information is easily available for decision makers and allows an easy parameterization of the model without resorting to an excessive number of parameters (which is one of the problems of ABM identified in the literature). These statistics are denoted as:

$$Q_{recovery}^h, Q_{worsening}^h, Q_{death}^h$$

defining the proportion of patients who eventually recovering/worsening/dying in state h. From this parameters, the remaining transition probabilities can be computed as:

$$P(worsening|h) = \frac{Q_{worsening}^h}{T_h}; P(recovery|h) = \frac{Q_{recovery}^h}{T_h}; P(death|h) = \frac{Q_{death}^h}{T_h}$$

considering that the overall probability of changing of state is $\frac{1}{T_h}$ and the aggregate proportions must be respected.

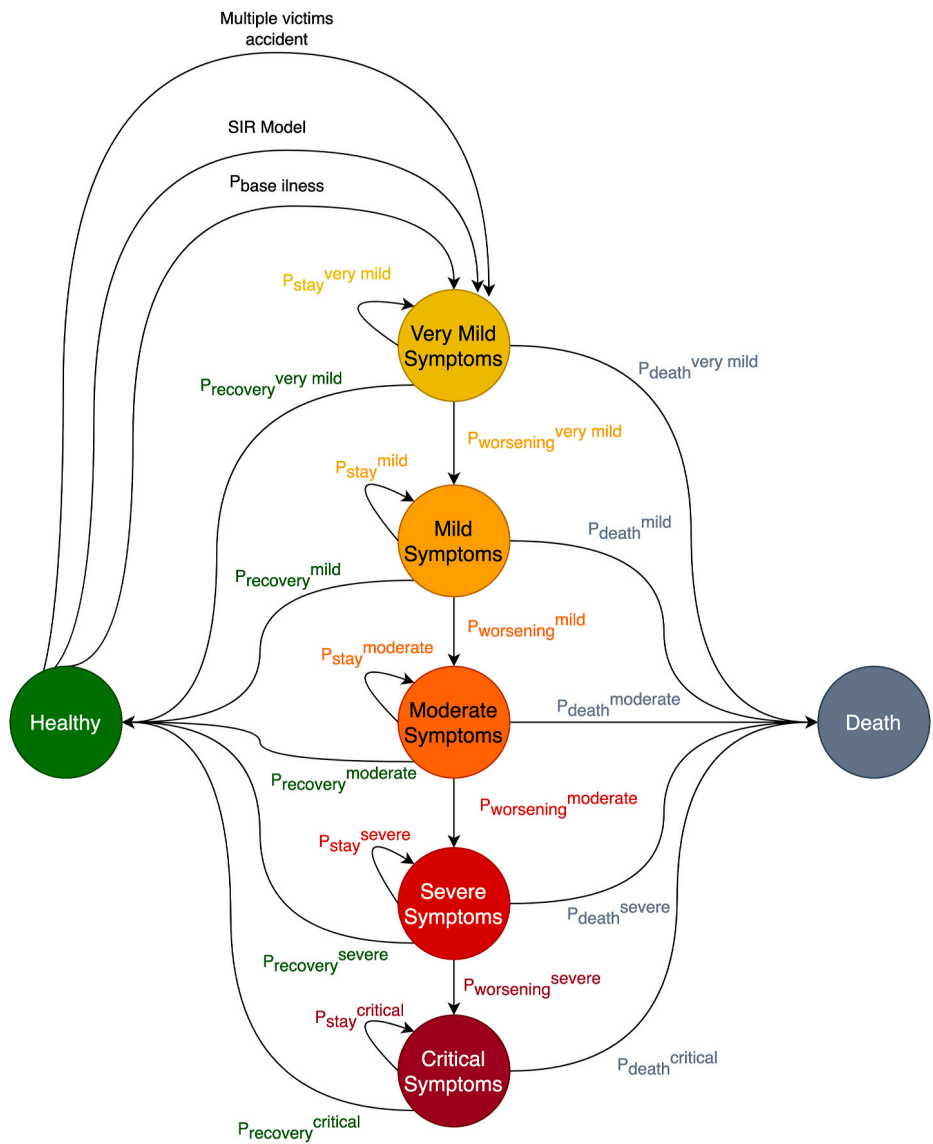

**Fig. 3.** Disease health evolution modelled as a markov chain.

Using the previous transition probabilities, it is possible to simulate over time the disease evolution of a Patient agent. Additionally, diseases with different severities (i.e.: different recovery times, proportion of patients reaching critical stages, different death rates …) can be modelled by tuning the aggregated parameters of different independent *Disease* (denoted *di*) instances (which may be simulated concurrently). That is, the required parameters to model a disease are:

$$di = \left(T_h, Q_{recovery}^h, Q_{worsening}^h, Q_{death}^h\right) \forall h \in H$$

Another key aspect of the proposed disease evolution model is that health states may be linked with a required level of healthcare service that favors patients' recovery. In this sense, Hospital agents are in charge of providing healthcare services to patients. Four types of healthcare resources are envisioned in this article (as depicted in Fig. 4, and denoted by set C: assistance through mobile health services (appropriate for mild patients), in-person consultation with a clinician (optimal for patients with moderate symptoms), general hospitalization (required by patients for severe symptoms), and critical care hospitalization (used for critical patients):

$$c \in \{mHealth, inPerson, generalBed, ICU\} \equiv C$$

As the health state of a patient deteriorates, it can seek medical attention in a **Hospital** agent. However, healthcare resources are limited (a maximum capacity is enforced at each care level) which may result in the patient receiving medical care at a lower service than expected or even having to wait to receive any medical care at all. At each simulation step, the allocation of medical resources to patients is reevaluated so that patients may be promoted to higher level services as resources become available, or as needed due to potential health deterioration. Ultimately, the hospital agent ensures the best possible patient care with constrained resources using a First Come-First Served approach and providing the highest quality healthcare service available and compatible with the current illness state of the Patient.

Received healthcare services affect patients' outcome, particularly when a patient cannot access the required care. Thus, the previous transition probabilities governing disease evolution should be conditional to the received healthcare services. To this regard, the previous disease model must be extended to depend on the type of received medical care:

$$\left(T_{h,c}, Q_{recovery}^{h,c}, Q_{worsening}^{h,c}, Q_{death}^{h,c}\right) \forall h \in H, c \in C$$

where *h* represents the health state and *c* the care level received. To do so, for each parameter, a matrix or table similar to Table 2 must be constructed to represent the double dependency (i.e.: current state and medical care). In it, the patient's prognosis will worsen as the care received moves away from the care required. Then, at each simulation step, the appropriate set of parameters is used to compute the transition probabilities, extending the double dependency (i.e.: health state and health level) into the simulation. As in the previous cases, this aggregate information can be obtained from health records.

Additionally, healthcare facilities Quality of Service (QoS) might not be constant as it can be dependent on hospital level of occupancy (i.e.: saturated services tend to perform worse) or other factors such as the dependence with the IC services. To model this, an *attention quality* level is introduced at each care service ($c \in C$). This parameter can be manually tuned or be dependent on some of the previous factors. As the attention quality declines, the prognosis of patients is expected to worsen. To introduce this dependency, it is necessary to establish a relationship between the attention quality and the transition probabilities that shape the course of the disease. In this sense, let us assume that the probabilities calculated so far correspond to those obtained for a quality of service equal to 1, i.e., the nominal behavior of each service. Similarly, let us assume that as the quality of care degrades, the transition probability is modified up to a limit (i.e.: non-nominal probability in Fig. 5) that will be matched to the probabilities that would be obtained at the previous level of care. That is, if a critical hospitalization service has an attention quality of 0, that would be equivalent to receiving the care of a general hospitalization bed. Then, at each simulation step, the transition probability is gradually changed (e.g.: following a quadratic law) between the nominal and non-nominal probabilities as modulated by the attention quality. This relationship is depicted in Fig. 5.

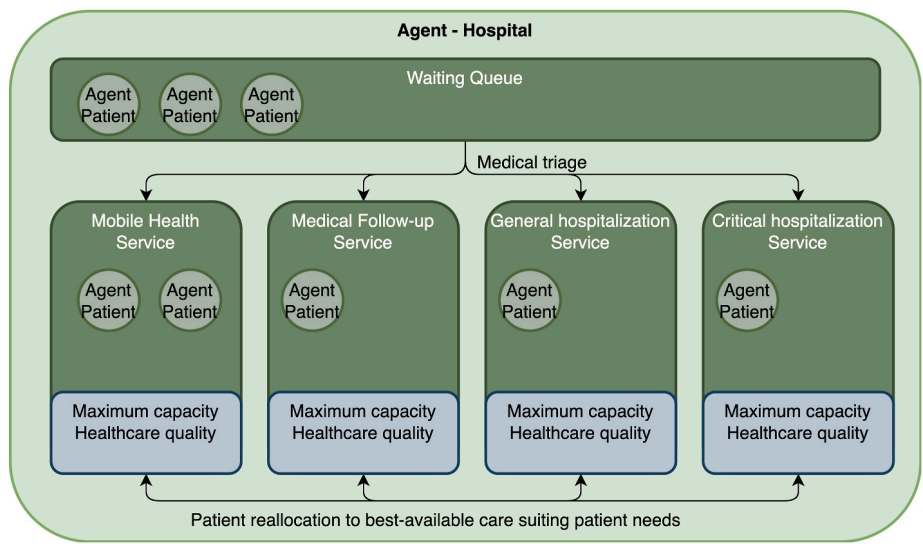

**Fig. 4.** Agent hospital model with different healthcare services.

**Table 2**
Table to define aggregated parameters to model disease evolution and its dependence with the received health care. As an example, the average length state in each state is depicted (T). Similar tables would be needed for the rest of aggregated parameters (Q).

| Medical care | Health state | | | | |
|---|---|---|---|---|---|
| | Very mild symptoms | Mild symptoms | Moderate symptoms | Severe symptoms | Critical symptoms |
| **No follow up** | $T_{very\ mild,\ no\ followup}$ | $T_{mild,\ no\ followup}$ | $T_{moderate,\ no\ followup}$ | $T_{severe,\ no\ followup}$ | $T_{critical,\ no\ followup}$ |
| **Mhealth** | – | $T_{mild,\ mhealth}$ | $T_{moderate,\ mhealth}$ | $T_{severe,\ mhealth}$ | $T_{critical,\ mhealth}$ |
| **In-person consultation** | – | – | $T_{moderate,\ in-person}$ | $T_{severe,\ in-person}$ | $T_{critical,\ in-person}$ |
| **General hospitalization** | – | – | – | $T_{severe,\ general}$ | $T_{critical,\ general}$ |
| **Critical hospitalization** | – | – | – | – | $T_{critical,\ icu}$ |

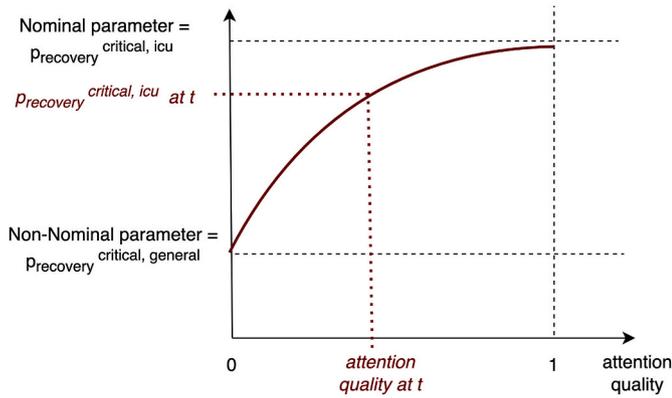

**Fig. 5.** Attention quality effect on recovery probability.

So far, the model has been detailed with a focus on disease progression. In turn, the relationship between agents is defined as a probabilistic dependency matrix that governs how patients from each population to a set of hospitals to which patients are sent. Moreover, hospitals behave as a networked healthcare provider, meaning that patients can be transferred between hospitals once their maximum capacity has been reached.

The previous model can be easily parameterized to represent different scenarios with an assorted number of agents, different diseases, or service capacities. Then, as an **output**, among other, the model provides information regarding *healthcare service usage* (available and occupied capacity per level), *unattended patients' statistics* (waiting times), treated *patients' statistics* (mean treatment time) *incidence of different diseases or deaths* in each population and *deaths* in each population.

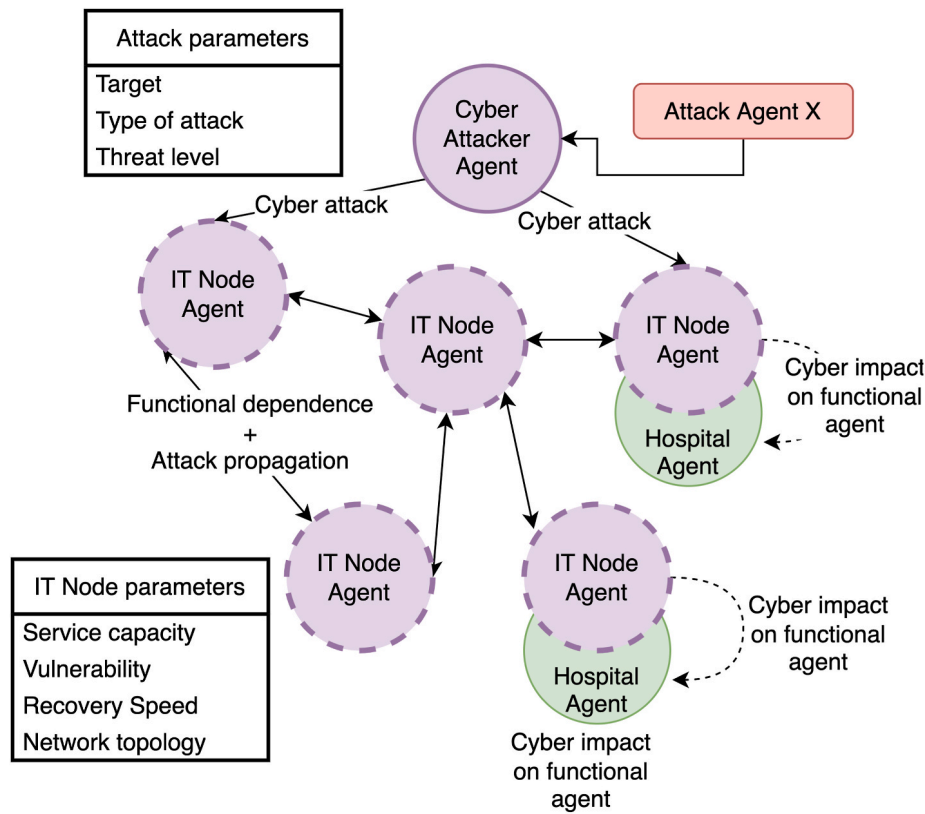

**Fig. 6.** ABM model of the IT system and interdependence with HS.

*3.3. Information & telecommunication system simulation*

An information and telecommunication (IT) system ABM is executed concurrently with the HS model to represent the information communication and processing capabilities needed for its operation. This system, in the current version of our simulation platform, is also modelled as a network of two types of autonomous agents (see Fig. 6): **cyber-attackers** and **IT nodes**.

The first type of agent are **cyber-attackers**, that generate attacks (with different parameterizable threat levels) against IT nodes. Different *types of attacks* can be modelled, i.e.

- Botnet (it infects other entities to be used to send ransomware or DDOS attacks),
- Ransomware (the infected node becomes unavailable for a given time) and
- DDOS (short-term attack which might yield an infrastructure unavailable depending on its capacity).

Attacks can be targeted to specific IT nodes or be general. Similarly, it is possible to model the difficulty in detecting the cyber-attack and the average response time to recover from the degradation caused by the attack.

The second one are **IT nodes themselves**, that represent functional entities (e.g.: servers, servers or IT service providers …) in a network that may be the subject of the cyberattacks. Each IT node agent can be parameterized with a different *service capacity* (number of requests that can be handled by unit of time), *vulnerability level* (classification of the probability to be affected by an incoming cyberattack) against attacks and a different *recovery capacity* (modulation over time required to recover from a suffered cyberattack). This way, it is possible to stochastically model cyberinfrastructure resiliency against an attack. If the attack is finally successful, the IT node might become degraded for some time.

Events (statically and programmatically defined) are used to model functional dependencies between IT nodes and the propagation of the cyber-attack processes. These dependencies are parameterized using a graph that represents the **network topology** in which the proper functioning of some IT nodes depends on one or more other agents (i.e.: other IT nodes). To model that the health care system depends on the underlying IT services, each hospital agent is coupled with an IT node agent operating within the IT system. If that IT agent becomes unavailable, the **interdependency generates a direct effect in the associated Health Agent**, which indirectly impacts the HS as a whole. Depending on the type of attack and unavailability time, hospital services might become unavailable (especially those reliant on the IT system such as the mHealth service) or service quality of service might be degraded (affecting patient outcome and increasing patient mean stay time in each service). As an added value **output**, the simulator allows to analyse the number of received attacks in each IT node and its availability/quality of service over time.

*3.4. Implementation remarks*

The proposed simulation model has been implemented using Mesa [37], an ABM framework for Python. This framework maps the ABM paradigm into object-oriented programming in which each type of agent corresponds to a class where attributes define the state of the agent, and functions define agents' interactions. It also provides scheduling utilities to implement a discrete time simulation, ensuring that the behaviours of all agents are run in every simulation step.

A file-based interface has been designed to configure simulation scenarios. It allows defining which agents to simulate along with their parameters, dependencies, and network topology. As an output, a CSV file is provided with the resulting timeseries and KPI. This file is easily processed in Python to provide an understandable, graphical representation of the simulation results. The simulator is computationally lightweight and parallelizable. Montecarlo experiments can also be run to provide statistically relevant results.

Regarding the ethical considerations of the implemented system, the European High-Level Expert Group (HLEG) on AI [38] has delivered a set of key ethical requirements that are relevant for the development and usage of Agent-Based Modelling. These requirements include *human agency and oversight*, which is guaranteed in the proposed system as it does not intend to replace a human decision-maker (DM). Instead, it aims to support in the decision-making process by providing valuable metrics which will guide the decision taken autonomously by a human DM. Another relevant requirement is *data privacy and governance*. To use the tool, DMs will need to abide to applicable regulation such as HIPPA [39] or GDPR [40], which is possible as only aggregated, anonymized data is needed to tune the proposed model. *Transparency and explainability* are guaranteed thanks to the dual micro-macro nature of ABM simulation. Thus, it possible to explain the high-level, aggregated results provided as an output to DMs from the low-level individual behaviour of each agent. In this regard, it is also essential to communicate to DMs the modelling assumptions and the limitations that they may entail. Furthermore, the principle of *technical robustness and safety* has been implemented by proposing a simulation model that extends and integrates state-of-the-art models (e.g.: SIR, Markov models for disease evolution) together with sensible hypothesis. Also, the model has been validated following a literature-supported methodology (as will be discussed in Section 4.1). Given the critical implications of decisions taken on HS, it may be appropriate to implement additional safeguards such as periodic validations or the use models ensembles. The principle of *diversity, non-discrimination and fairness* ensures that models are inclusive and avoid bias. This may be a limitation of the proposed model as it does not allow to study separately different population groups (e.g.: age, gender) to assess whether there are impacted differently by threats or countermeasures. *Accountability* shall be achieved through audits within the organization where the decision-making tool is implemented. Moreover, peer-review and open-access of this article also contributes to this aspect. Finally, the objective of the proposed tool is to *improve societal well-being*, the last of the HLEG principles.

**4. Validation**

*4.1. Validation methodology*

As already identified in Section 2.2, validating ABM is a challenging endeavour due to the parameterization dimensionality and usual stochasticity of agent behaviour. In fact, there does not exist a broadly accepted method to perform ABM validation. Authors of [41] propose a 4 step approach to validate and test an ABM:

1. *Cross-validation.* If the model has been generated as an extension of a well-known model or if it exits an already validated model for the same problem, the outputs of the ABM can be compared with the ones of the previously simulated model.
2. *Sensitivity analysis.* This step consists in assessing if the output of the model changes as expected to variations of the input parameters.
3. *Comparison to data.* If available, data extracted from the real system is a valuable tool to validate the model. However, in many processes real data might not be available or be subjected to access restrictions.
4. *Model testing.* As a last step, the usefulness of the proposed model must be assessed, ensuring that it can provide new, valid insights for a decision-maker in a real scenario.

Although all steps should be ideally performed in a new model,

authors argue that due to the complexity of ABM and possible limitations on comparable information, it might not be possible to carry out all steps. Regarding steps 1 to 3, several technical methods can be used to carry out them. As discussed in Refs. [42,43], one can resort to *data-driven/statistical methods* to assess the obtained results. A proper statistical analysis of the model might require a significative number of runs covering an extensive combination of input parameters to provide. Alternatively, *face validation or visualization* techniques can be also used to either compare different models' outputs (i.e.: step 1), verify that the model responds to changes in the input data (i.e.: step 2) or compare the results with real data (i.e.: step 3). Face validation consists in a human expert assessing that the simulated system correctly imitates the real system, and its result are coherent. This can be realized using graphical representations of the model output and agent's internal state over time.

Based on the previous methodology, we propose to articulate the validation steps through a use case that will be described in Section 4.2. The use case will present a representative usage scenario of the simulation tool as a DSS. As an output of the use case execution, agent's behaviour over time and aggregate statistics will be obtained. With them, the following validation steps will be performed using the face validation technique.

- Step 1. A similar model does not exist in the literature. Thus, it is not possible to directly perform cross-validation. However, the proposed model builds upon the SIR disease model. Thus, the obtained disease propagation curves must be consistent with the typical "bell-like" graph of SIR models.
- Step 2. The execution of the use case involves running the simulation model with a matrix of different combinations of the configuration parameters (e.g., different threats, different contingency measures). This serves as a sensitivity analysis as it allows to check that varying the input data leads to different results, and that these results are as expected. For example, an increase in the infection rate should lead to a higher number of infections and consequently a higher number of hospitalizations.
- Step 3. Unfortunately, hospital occupancy, wait times or cyber-attacks information is usually not openly available to the public. Thus, it is not possible to perform a comparison with real data or perform statistical validation.
- Step 4. The use case will prove that the proposed models can generate useful insights to the decision maker who will be able to compare different contingency responses to enhance resiliency through a contingency matrix.

### 4.2. Validation use-case description

As explained in the previous subsection, the validation use-case is the guiding thread to validate the model. Through its execution, cross-validation, sensitivity analysis and model testing will be performed. Moreover, it will demonstrate the feasibility and flexibility of the proposed simulation tool for decision making in resiliency enhancing scenarios with uncertainty and for comparing different decision alternatives.

Let us imagine that a regional-level healthcare provider wants to improve the resiliency of its operation against compounding-mixed threats. To do so, the healthcare provider models its infrastructure as a set of autonomous agents following the proposed architecture, as shown in Fig. 7. Two towns A and B (of 40,000 and 150,000 inhabitants, respectively) are served by two homonymous hospitals with on-site consultation services, general hospitalization, and critical care units (each with different maximum capacities). In particular, the decision maker wants to evaluate and improve the resilience of Hospital B, with an in-person attention capacity of 1000 people, 278 beds and 30 critical beds respectively. The hospital has no dedicated cyber defence resources, so it is very vulnerable to potential cyber-attacks. Additionally, in case of need, hospital A, which has a larger capacity, can support the needs of hospital B by accepting referred patients.

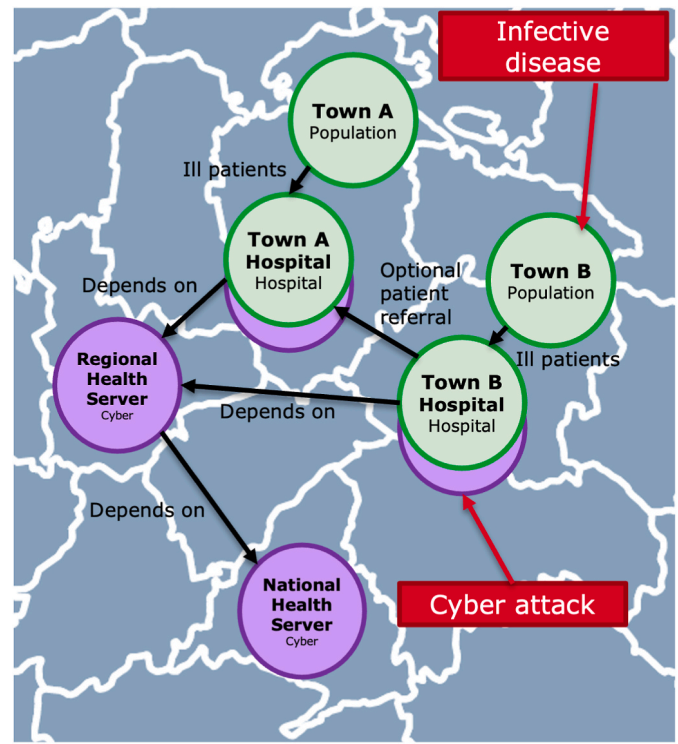

**Fig. 7.** Validation scenario. The IT system operates at hierarchically dependent levels: national level, regional level, and hospital level. The health care system is composed of two hospitals in two different populations: Town A and Town B. Town A hospital may support Town B hospital if needed, accepting transferred patients.

The decision-maker wants to consider the coexistence of a cyber-threat against Hospital B information & telecommunications system and the spread of an infective disease within Population B. These compounding threats can occur with different intensity levels, with the impact on the infrastructure varying accordingly. Therefore, it is of interest to evaluate a set of scenarios that consider risks' variability and unpredictability. In this regard, the ABM simulator makes it possible to easily parameterize different risk scenarios and evaluate the impact on the hospital under analysis. In this way, a baseline scenario is configured consisting of the 8 possible combinations of the following parameters.

- A cyber-attack with high severity (i.e.: high recovery time, denoted *highAttack*) and a cyberattack with low severity (named *lowAttack*).
- An infectious disease with a high contagion rate (i.e.: fast spread over the population, referred as *highContagion*) and a disease with a lower contagion rate (denoted as *lowContagion*).
- The previous infectious disease having high pathogenicity (i.e.: it causes high severe disease with high mortality and requiring advanced healthcare, named *highSeverity*) and another having lower severity (referred as *lowSeverity*).

These risk scenarios (i.e.: a combination of *highAttack/lowAttack*, *highContagion/LowContagion*, *highSeverity/lowSeverity*) can be easily parameterized within the proposed ABM model. The required parameters (already described in Section 3) are aggregated parameters that are usually accessible by decision-makers or healthcare managers. For instance, the pathogenicity level of a disease can be set by tuning the $Q_{death}^h$ parameter at each health state and severity level as shown in Table 3 below. Similarly, the disease contagion level can be set tuning the transmission rate of the SIR model.

A Montecarlo simulation with ten realizations of each combination has been performed to obtain valuable metrics to assess the impact of each of the compounding threats combination. Moreover, this parameter variation will serve as the basis for the sensitivity analysis of the model. Each realization of the simulated threat scenarios will simulate the HS and supporting IT system for 350 days.

**Table 3**
Parameterized death aggregate rate ($Q^h_{death}$) at each level and health state, in number of deaths per 100 patients. Yellow cells show the values for the *lowSeverity* scenarios whereas red cells show the used values in the *highSeverity* scenario.

| Medical care \ Health state | Very mild symptoms | | Mild symptoms | | Moderate symptoms | | Severe symptoms | | Critical symptoms | |
|---|---|---|---|---|---|---|---|---|---|---|
| No followup | 0.05 | 0.05 | 0.05 | 0.05 | 2 | 2 | 30 | 25 | 99 | 99 |
| Mhealth | - | - | 0.05 | 0.05 | 0.05 | 0.05 | 15 | 20 | 80 | 95 |
| In-person consultation | - | - | - | - | 0.05 | 0.05 | 10 | 10 | 40 | 80 |
| General hospitalization | - | - | - | - | - | - | 2 | 2 | 30 | 40 |
| Critical hospitalization | - | - | - | - | - | - | - | - | 10 | 20 |

### 4.3. Use case part 1: analysing compounding threats impact on the HS

After running the scenarios, visualization techniques can be used to assess the impact on the HS of each risk scenario. Moreover, they will also serve to perform cross-validation and sensitivity analysis using a face validation technique. In this sense, Figs. 8–10 show the average results (across the 10 Montecarlo runs) for a subset of the analysed risk scenarios. For clarity, the figures show only a subset of the risk combinations.

Regarding the materialized threats, Fig. 8(a) shows the evolution of the number of people infected by the simulated contagious disease. As expected in a transmission model derived from a SIR model, the disease incidence follows a "bell" curve (validation step 1). Also, it can be seen how the model correctly responds to changes in the configured parameters. In this sense, high levels of infectivity (*highContagion* scenarios) lead to a high peak of disease incidence (blue and green curves) or, on the contrary (*lowContagion* scenarios), to a flattened curve (orange and red curves). The cyberattack risk is displayed in Fig. 8(b) that shows how the occurrence of a cyberattack degrades the hospital's information services. Again, the different configured levels of cyberattack severity result in different service unavailability times. A low severity cyberattack (*lowAttack* scenarios) can be resolved in a shorter time (green and red curves), whereas a high severity one (*highAttack* scenarios) requires a substantial recovery time (blue and orange curves). Therefore, this proves that the simulation model is adequately sensitive to changes in the input parameters (validation step 2).

The final impact of these threats on the population at a macro level is shown in Fig. 8(c) that depicts the average number of reported deaths. Further supporting the sensitivity analysis, it can be seen how the materialization of both threats can create a surge in deaths that is temporally aligned with the occurrence of those threats. This is clearly visible in the orange curve with a first surge (around t = 100) corresponds with the cyberattack effect and a second surge (around t = 200) corresponds with the disease infections impact.

At a micro/agent level, the impact of these threats on hospital infrastructure can also be analysed, as shown in Fig. 9 which depicts the occupancy of the different health services of Hospital B (i.e.: in person consultation, general hospitalization and ICU hospitalization) across time. Again, the impact of the described threats depends on the experienced combination of parameters (risk scenarios, shown as different colour curves). A severe cyber-attack and a rapidly transmitted pathogenic pandemic (i.e.: blue curves in Fig. 9) cause a synergistic negative effect that is observed as a single peak in occupancy across all services. On the other hand, a slower-transmission pandemic (i.e.: orange and red curves in Fig. 9 depicts), can decouple the negative effects of the two threats by delaying the contagions peak. In both cases, the cyber-attack causes a first drop in the quality of care, which results in longer care times. This leads to an increase in hospital occupancy at all care levels. On the other hand, the pandemic causes an increase in hospital demand that may end up collapsing the hospital, as shown by the plateaus in occupancy once the maximum capacity has been reached. Different disease parameterization (i.e.: different Q values) may result in different rates of use of different services.

Once the collapse of a service has been reached, the collapse severity can be assessed by analysing the utilization rate (i.e.: ratio between the patient arrival rate and the mean occupancy rate) in Fig. 10. Utilization rates lower than 1 indicate that the service can meet the care demand. Then, after the 1-threshold is exceeded, the higher the utilization rate, the more degraded a health care service is, resulting in longer waiting times (patients are left unattended) and worse attention quality for attended patients. Degradation in hospital services ultimately impacts on the number of reported deaths (Fig. 8(c)) which is clearly superior in the most critical combination of threats (blue curve) versus the least threatening combination (red curve).

In the previous discussion, the model has been validated by cross-validation and sensitivity analysis. Moreover, the simulated use-case shows how the proposed simulation framework can assess the impact of varied cyber and natural threats in a HS and its associated populations, providing useful information to decision makers. A correct parameterization of the scenario allows to analyse the different impact of diseases with different transmission and medical care characteristics.

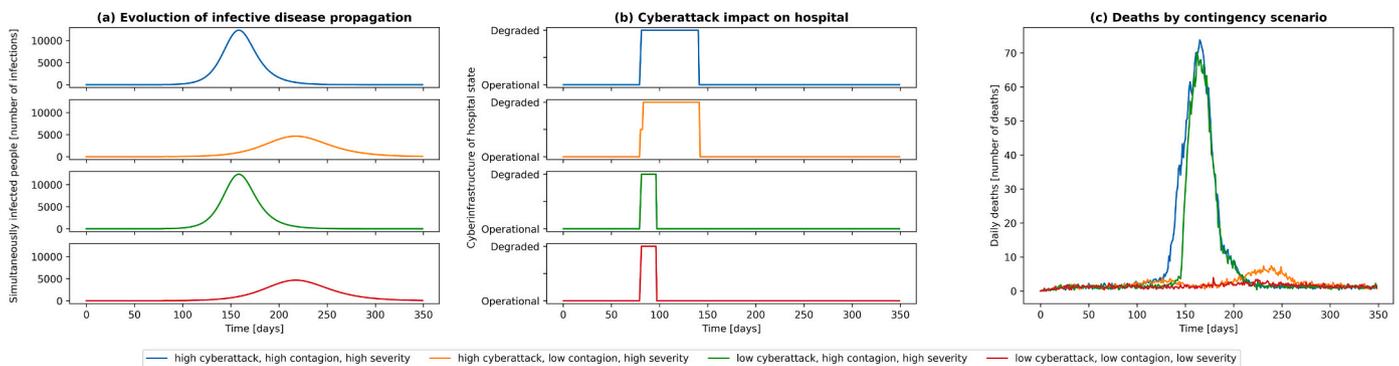

**Fig. 8.** (a) Mean evolution of contagious disease, (b) cyberattack impact for different severity levels, and (c) mean impact in terms of daily deaths. In all subfigures a subset of risk scenarios is shown as indicated by the color legend. (For interpretation of the references to colour in this figure legend, the reader is referred to the Web version of this article.)

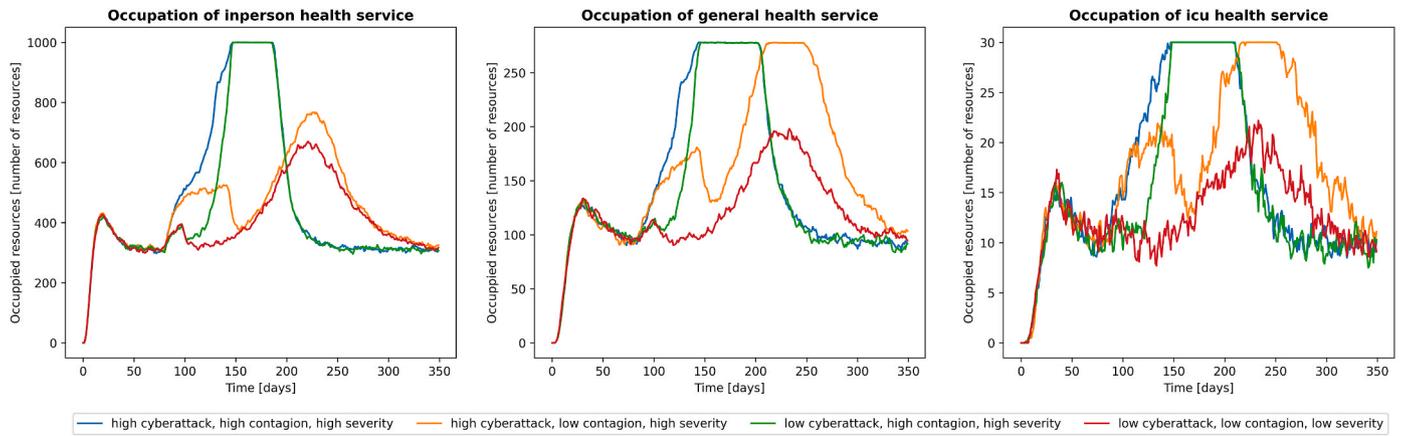

**Fig. 9.** Hospital B health services occupancy for different threat levels.

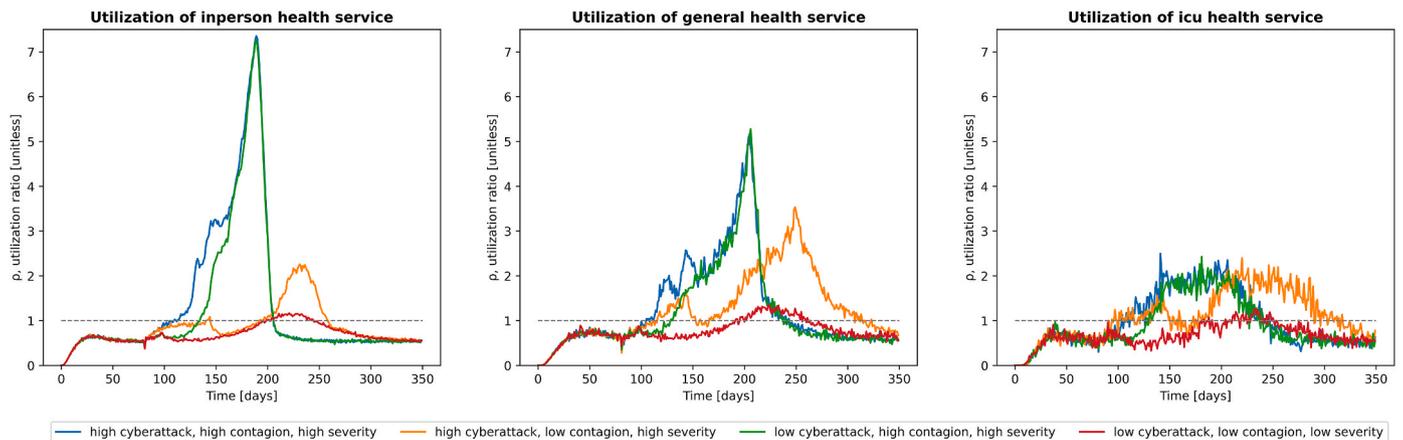

**Fig. 10.** Utilization rate (ratio between arrival rate and occupancy rate) of Hospital B services for the different threat levels.

*4.4. Use case part 2: making informed decisions to enhance resiliency*

The resulting number of deaths in some of the previous risk scenarios (as depicted in Fig. 8(c)) may not be deemed admissible. In that case, the decision maker must enhance the resilience of hospital B to reduce the experienced overload and the cost in terms of deaths and/or money. Consequently, after a detailed analysis of the impact results (with the above metrics or other available metrics such as waiting times per health care service, death rates for each service …), different alternative decisions/countermeasures might be proposed possible within the limited available resources. Specifically, in this example use case, the following contingency measures are considered.

- An increase of 50 % (*lowBeds* alternative) or 100 % (*highBeds* alternative) in the number of available resources in some healthcare services.
- A dedicated cyber-defense capacity that increases the hospital's *recovery capacity* and that it is expected to reduce the recovery time after the cyber-attack to 15 days (*lowSecurity* alternative) or 24 h (*highSecurity* alternative).
- The inclusion of a remote health monitoring service (*mHealth* alternative) for home monitoring of patients.
- A networked operation of the healthcare system, referring exceeding patients of Hospital B to Hospital A (*referral* alternative).

The optimal decision alternative may not be the same for all scenarios. In this sense, the proposed ABM simulator can be used again to evaluate the obtained improvement with each of the potential decision alternatives. Thus, for each of the 8 risk scenarios, a set of contingency scenarios is run combining the threats with each of the proposed contingency measures. To analyse the results in an orderly manner, contingency matrices may be used. These matrices are double-entry tables that show (for each decision alternative and each possible risk scenario) a merit function to assess its suitability.

After running the simulations, Figs. 11 and 12 show two contingency matrices using respectively the number of deaths and the utilization rate of medical services as figure of merit. Rows correspond to each of the risk scenarios detailed in Section 4.2, whereas columns correspond to each of the contingency/response options. An additional column named "baseline" corresponds to the reference result when no measures are implemented. Also, a colour code across each row is used to easily compare alternatives from the ones obtaining the best figure (green) to the ones obtaining the worst figure of merit value (red).

From the contingency tables, a decision maker could conclude that in cases with low severity contingencies (e.g.: low attack severity, low contagion rate and low disease severity) the proposed alternatives hardly reduce the number of reported deaths, although they are capable, in some cases, of reducing the system utilization rate. However, the number of deaths can be significantly reduced in the case of high severity contingencies. Especially, with patient referral alternatives or with a high increase in the number of beds. The mHealth alternative provides intermediate results that may be interesting since the implementation cost of an mHealth service may be lower than that of the other alternatives. On the other hand, cybersecurity strategies only seem relevant in the case of low virulence and low contagion (i.e.: no compound threat).

|  | baseline | highBeds | highSecurity | lowBeds | lowSecurity | mHealth | referral |
|---|---|---|---|---|---|---|---|
| highAttack_highContagious_highSeverity | 3047.0 | 1730.0 | 2548.0 | 2344.0 | 2536.0 | 2092.0 | 1219.0 |
| highAttack_highContagious_lowSeverity | 1058.0 | 620.0 | 882.0 | 654.0 | 920.0 | 708.0 | 736.0 |
| highAttack_lowContagious_highSeverity | 785.0 | 607.0 | 664.0 | 612.0 | 702.0 | 666.0 | 696.0 |
| highAttack_lowContagious_lowSeverity | 552.0 | 558.0 | 497.0 | 550.0 | 508.0 | 568.0 | 541.0 |
| lowAttack_highContagious_highSeverity | 2461.0 | 1452.0 | 2528.0 | 1885.0 | 2505.0 | 1793.0 | 1119.0 |
| lowAttack_highContagious_lowSeverity | 906.0 | 566.0 | 854.0 | 606.0 | 893.0 | 620.0 | 644.0 |
| lowAttack_lowContagious_highSeverity | 691.0 | 564.0 | 698.0 | 582.0 | 708.0 | 654.0 | 621.0 |
| lowAttack_lowContagious_lowSeverity | 501.0 | 490.0 | 503.0 | 495.0 | 499.0 | 494.0 | 505.0 |

**Fig. 11.** Contingency matrix taking the accumulated number of deaths as the merit function of each decision alternative. The baseline column shows the number of deaths in the baseline scenario. In the other columns, a lower value indicates that the alternative succeeds in reducing the impact of the threat.

|  | baseline | highBeds | highSecurity | lowBeds | lowSecurity | mHealth | referral |
|---|---|---|---|---|---|---|---|
| highAttack_highContagious_highSeverity | 5.15 | 4.06 | 5.54 | 4.66 | 5.42 | 6.56 | 2.36 |
| highAttack_highContagious_lowSeverity | 2.82 | 1.39 | 2.52 | 1.91 | 2.48 | 2.77 | 2.14 |
| highAttack_lowContagious_highSeverity | 3.5 | 1.25 | 3.92 | 1.66 | 3.57 | 3.6 | 2.14 |
| highAttack_lowContagious_lowSeverity | 1.53 | 0.81 | 1.36 | 0.91 | 1.36 | 1.71 | 1.43 |
| lowAttack_highContagious_highSeverity | 5.28 | 3.8 | 5.87 | 4.61 | 5.79 | 6.75 | 2.36 |
| lowAttack_highContagious_lowSeverity | 2.66 | 1.3 | 2.43 | 1.81 | 2.7 | 2.36 | 1.93 |
| lowAttack_lowContagious_highSeverity | 3.38 | 1.23 | 3.84 | 1.61 | 3.97 | 3.27 | 2.18 |
| lowAttack_lowContagious_lowSeverity | 1.34 | 0.67 | 1.34 | 0.92 | 1.33 | 1.23 | 1.36 |

**Fig. 12.** Contingency matrix taking the maximum utilization rate as the merit function of each decision alternative. The baseline column shows the number of deaths in the baseline scenario. In the other columns, a lower value indicates that the alternative succeeds in reducing the impact of the threat.

Overall, these contingency tables show how the simulation tool enables the decision maker to compare decision alternatives considering interdependences between healthcare and IT systems. Also, these tables could be enhanced by combining several decision alternatives (which can be simulated with the proposed tool) to check for synergic effects. These tables can be understood in two ways. On the one hand, from a strategic point of view, they can allow the decision-maker to select the best investment alternative that globally maximizes the merit function by giving a probability to each of the contingency scenarios. On the other hand, from a tactical point of view, they constitute an action guide for the decision-maker who, in the event of encountering one of the described risk scenarios, can easily select the most appropriate strategy. In both cases, they prove that the simulation framework is a valuable tool for the decision maker.

## 5. Conclusions

Health systems are subject to multiple risks that can degrade their performance level. These infrastructures are critical for societies and their resilience must be ensured. Unfortunately, the intrinsic complexity and interdependencies of these systems make them difficult to understand and predict, thereby hindering the ability to make informed decisions in response to risks. Reviewed literature shows that resilience frameworks enabling HS damage assessment, and system restoration alternatives comparison while encompassing interdependences are scarce. As a possible solution, simulator-based decision support systems can help to generate reliable metrics that assess systems in hypothetical scenarios to analyse risks.

In this sense, this article has proposed an ABM model aimed at simulating and analysing a healthcare system to facilitate evidence-based decision-making. The system is composed of a set of networked hospitals providing medical services at different levels: mHealth services, in-person consultations, general and critical hospitalization. In turn, demand can be simulated as a combination of a base demand, the spread of contagious diseases and the occurrence of multi-casualty accidents. These diseases are simulated at the patient level by means of a parameterizable Markov chain that can lead to different disease severities. The impact of hospital care on recovery has been addressed by considering the effect of the different care levels, waiting times and queues, and a possible reduction in hospital quality depending on the available resources and stress situation. Moreover, as a novelty, dependencies with information and telecommunication systems (IT) are considered within the proposed model through the simulation of a set of superimposed IT nodes (e.g.: servers, routers …) agents. These agents can be affected by a set of cyber-attacks that degrade their level of service, directly impacting hospitals services' ability to care for patients or even halting the provision of some kinds of care (i.e. mHealth).

The usefulness of the tool to measure the impact of different contingencies in the healthcare system has been demonstrated by means of a representative use case. Likewise, it has been explained how the tool can inform tactical and strategic resiliency-enhancing decision making through the generation of contingency matrices. Also, the ethical implications of the proposed tools have been discussed using the HLEG framework. In this sense, the tool ensures DM autonomy while providing explainable and robust results that may help HS resiliency enhancing. As a limitation, the presented model is a proof of concept that has been evaluated by face-validation [43] in a set of simplified scenarios (e.g.: just two hospitals and populations). Further statistical validation on more complex scenarios would benefit the accuracy of the results. Also, the modelling process necessarily involves the loss of some subtleties that will be noticed during the actual use of the tool (e.g.: the need to consider new hospital services or more complex disease transmission models). In this sense, the ABM paradigm allows to extend the proposed model in a simple way by refining the models of existing agents or by proposing new agents.

On future work, the model can be extended to include interactions with other critical infrastructures that affect the functioning of the health system: electrical system, goods distribution networks … Similarly, the use of explainable and causal artificial intelligence techniques may be explored to bridge the gap between the micro and macro levels of ABM. That would mean to automatically track system level anomalies or deviations to the individual behaviour of some agents. Also, the system would benefit from including advanced intelligent algorithm to automatically propose to the DM decision alternatives/countermeasures that help increase the resilience of the healthcare system. The model can be further enhanced to handle real-time events and dynamic analysis, evolving into a system capable of incorporating time-dependent decision analysis. This expansion enables not only the assessment of decision suitability but also the determination of the optimal timing for implementation. Moreover, specific data-driven aids to streamline model parametrization can be designed, so the model can be directly informed by real data. Ideally, this proposal of ABM should be integrated within a comprehensive simulation and analysis suite that provides a useful tool for holistic decision making.

**CRediT authorship contribution statement**

**David Carramiñana:** Writing – original draft, Validation, Software, Methodology, Investigation, Formal analysis, Conceptualization. **Ana M. Bernardos:** Writing – review & editing, Validation, Project administration, Methodology, Investigation, Conceptualization. **Juan A. Besada:** Writing – review & editing, Supervision, Methodology, Conceptualization. **José R. Casar:** Writing – review & editing, Supervision, Methodology, Funding acquisition, Conceptualization.

**Declaration of competing interest**

The authors declare that they have no known competing financial interests or personal relationships that could have appeared to influence the work reported in this paper.

**Acknowledgment**

Authors acknowledge sponsorship from Horizon Europe EDF project 101103386 'and Universidad Politécnica de Madrid project RP220022063. This research has been partially funded by Horizon Europe EDF project 101103386, by Spanish Ministerio de Ciencia e Innovación (MCIN/ AEI /10.13039/501100011033) project PID2020-118249RB-C21 and by Universidad Politécnica de Madrid project RP220022063.

**Data availability**

No data was used for the research described in the article.